\documentstyle[12pt]{article}

\begin{document}

\baselineskip=14pt plus 1pt minus 1pt

\begin{center}{\large \bf A sequence of potentials lying between
the U(5) and X(5) symmetries} 

\bigskip\bigskip

{Dennis Bonatsos$^{\#}$\footnote{e-mail: bonat@inp.demokritos.gr},
D. Lenis$^{\#}$\footnote{e-mail: lenis@inp.demokritos.gr}, 
N. Minkov$^\dagger$\footnote{e-mail: nminkov@inrne.bas.bg}, 
P. P. Raychev$^\dagger$\footnote{e-mail: raychev@phys.uni-sofia.bg, 
raychev@inrne.bas.bg},
P. A. Terziev$^\dagger$\footnote{e-mail: terziev@inrne.bas.bg} }
\bigskip

{$^{\#}$ Institute of Nuclear Physics, N.C.S.R.
``Demokritos''}

{GR-15310 Aghia Paraskevi, Attiki, Greece}

{$^\dagger$ Institute for Nuclear Research and Nuclear Energy, Bulgarian
Academy of Sciences }

{72 Tzarigrad Road, BG-1784 Sofia, Bulgaria}

\end{center}

\bigskip\bigskip
\centerline{\bf Abstract} \medskip
Starting from the original collective Hamiltonian of Bohr and separating the 
$\beta$ and $\gamma$ variables as in the X(5) model of Iachello, 
an exactly soluble model corresponding to a harmonic oscillator potential 
in the $\beta$-variable (to be called X(5)-$\beta^2$) is constructed.
Furthermore, it is proved that the potentials of the form $\beta^{2n}$ 
(with $n$ being integer) provide a ``bridge'' between this new X(5)-$\beta^2$
model (occuring for $n=1$) and the X(5) model (corresponding to an infinite 
well potential in the $\beta$-variable, materialized for $n\to \infty$). 
Parameter-free (up to overall 
scale factors) predictions for spectra and B(E2) transition rates are 
given for the potentials $\beta^2$, $\beta^4$, $\beta^6$, $\beta^8$, 
corresponding to $R_4=E(4)/E(2)$ ratios of  2.646, 2.769, 2.824, and 2.852 
respectively, compared 
to the $R_4$ ratios of 2.000 for U(5) and 2.904 for X(5). Hints about nuclei 
showing this behaviour, as well as about potentials ``bridging'' the X(5) 
symmetry with SU(3) are briefly discussed. 
 


{\bf 1. Introduction} 

Models providing parameter-independent predictions for nuclear spectra 
and electromagnetic transition rates serve as useful benchmarks in nuclear 
theory. The recently introduced E(5) \cite{IacE5} and X(5) \cite{IacX5}  models
belong to this category, since their predictions for nuclear spectra 
(normalized to the excitation energy of the first excited state) and B(E2) 
transition rates (normalized to the B(E2) transition rate
connecting the first excited state 
to the ground state) do not contain any free parameters. The E(5) model 
appears to be related to a phase transition from U(5) (vibrational) to O(6) 
($\gamma$-unstable) nuclei \cite{IacE5}, while X(5) is related to a phase 
transition from U(5) (vibrational) to SU(3) (prolate deformed) nuclei 
\cite{IacX5}. Both models 
originate (under certain simplifying assumptions) 
from the Bohr collective Hamiltonian \cite{Bohr}, which is known to possess 
the U(5) symmetry of the five-dimensional (5-D) harmonic oscillator 
\cite{CM870}. 

In the present paper we study a sequence of potentials lying 
between the U(5) symmetry of the Bohr Hamiltonian and the X(5) model. 
The potentials, which are of the form $u_{2n}(\beta) = \beta^{2n}/2$, 
with $n$ being integer, are depicted in Fig. 1.  
For $n=1$ an exactly soluble model with $R_4=E(4)/E(2)$ ratio 
equal to 2.646 is obtained, while X(5) (which corresponds to an infinite well 
potential) occurs for $n\to \infty$ 
(in practice $n=4$ is already quite close to X(5)).  
Parameter-independent predictions for the spectra and B(E2) values 
(up to the overall scales mentioned above) are obtained for 
the potentials $\beta^2$, $\beta^4$, $\beta^6$, $\beta^8$. 
In addition to providing a number of models giving predictions directly 
comparable to experiment, the present sequence of potentials shows the way 
for approaching the X(5) symmetry from the direction of U(5) and gives a hint 
on how to approach the X(5) symmetry starting from SU(3). 

In Section 2 of the present paper the exactly soluble model obtained 
with the $\beta^2$ potential, to be called X(5)-$\beta^2$, is introduced 
and compared to X(5), while in Section 3 a sequence of potentials lying 
between the X(5)-$\beta^2$ and X(5) models is considered.
Numerical results for spectra and B(E2) transition rates are given for all
these potentials, which lie between the U(5) symmetry of the Bohr 
Hamiltonian \cite{Bohr,CM870} and the X(5) model. A brief comparison 
to experimental data is given in Section 4, while in Section 5  
perspectives for further theoretical work are 
discussed and the conclusions are summarized.  

{\bf 2. X(5)-$\beta^2$: A new exactly soluble model} 

{\bf 2.1 The $\beta$-part of the spectrum} 

The original Bohr Hamiltonian \cite{Bohr} is
\begin{equation}\label{eq:e1}
H = -{\hbar^2 \over 2B} \left[ {1\over \beta^4} {\partial \over \partial 
\beta} \beta^4 {\partial \over \partial \beta} + {1\over \beta^2 \sin 
3\gamma} {\partial \over \partial \gamma} \sin 3 \gamma {\partial \over 
\partial \gamma} - {1\over 4 \beta^2} \sum_{k=1,2,3} {Q_k^2 \over \sin^2 
\left(\gamma - {2\over 3} \pi k\right) } \right] +V(\beta,\gamma),
\end{equation}
where $\beta$ and $\gamma$ are the usual collective coordinates, while
$Q_k$ ($k=1$, 2, 3) are the components of angular momentum and $B$ is the 
mass parameter.  

One seeks solutions of the relevant Schr\"odinger equation having 
the form 
$ \Psi(\beta, \gamma, \theta_i)= \phi_K^L(\beta,\gamma) 
{\cal D}_{M,K}^L(\theta_i)$, 
where $\theta_i$ ($i=1$, 2, 3) are the Euler angles, ${\cal D}(\theta_i)$
denote Wigner functions of them, $L$ are the eigenvalues of angular momentum, 
while $M$ and $K$ are the eigenvalues of the projections of angular 
momentum on the laboratory-fixed $z$-axis and the body-fixed $z'$-axis 
respectively. 

As pointed out in Ref. \cite{IacX5}, in the case in which the potential 
has a minimum around $\gamma =0$ one can write  the last term of Eq. 
(\ref{eq:e1}) in the form 
\begin{equation}\label{eq:e3} 
\sum _{k=1,2,3} {Q_k^2 \over \sin^2 \left( \gamma -{2\pi \over 3} k\right)}
\approx {4\over 3} (Q_1^2+Q_2^2+Q_3^2) +Q_3^2 \left( {1\over \sin^2\gamma}
-{4\over 3}\right).  
\end{equation}
Using this result in the Schr\"odinger equation corresponding to 
the Hamiltonian of Eq. (\ref{eq:e1}), introducing reduced energies 
 $\epsilon = 2B E /\hbar^2$ and reduced potentials $u = 2B V /\hbar^2$,  
and assuming that the reduced potential can be separated into two terms, 
one depending on $\beta$ and the other depending on $\gamma$, i.e. 
$u(\beta, \gamma) = u(\beta) + u(\gamma)$, the Schr\"odinger equation can 
be separated into two equations \cite{IacX5}  
\begin{equation} \label{eq:e5}
\left[ -{1\over \beta^4} {\partial \over \partial \beta} \beta^4 
{\partial \over \partial \beta} + {1\over 4 \beta^2} {4\over 3} 
L(L+1) +u(\beta) \right] \xi_L(\beta) =\epsilon_\beta  \xi_L(\beta), 
\end{equation}
\begin{equation}\label{eq:e6} 
\left[ -{1\over \langle \beta^2\rangle \sin 3\gamma} {\partial \over 
\partial \gamma}\sin 3\gamma {\partial \over \partial \gamma} 
+ {1\over 4 \langle \beta^2 \rangle } K^2 \left( {1\over \sin^2 \gamma}
-{4\over 3}\right) +u(\gamma)\right] \eta_K(\gamma) = 
\epsilon(\gamma) \eta_K(\gamma),
\end{equation}
where $\langle \beta^2 \rangle$ is the average of $\beta^2$ over $\xi(\beta)$ 
and $\epsilon= \epsilon_\beta +\epsilon_\gamma$. 
 
In Ref. \cite{IacX5} Eq. (\ref{eq:e5}) is solved exactly for the case 
in which $u(\beta)$ is an infinite well potential
\begin{equation}\label{eq:e7} 
 u(\beta) = \left\{ \begin{array}{ll} 0 & \mbox{if $\beta \leq \beta_W$} \\
\infty  & \mbox{for $\beta > \beta_W$} \end{array} \right. .  
\end{equation} 
The relevant exactly soluble model 
is labeled as X(5) (which is not meant as a group label, although 
there is relation to projective representations of E(5), the Euclidean 
group in 5 dimensions \cite{IacX5}). In particular, Eq. (\ref{eq:e5}) in the 
case of 
$u(\beta)$ being an infinite well potential is transformed into a Bessel 
equation, the relevant eigenvalues being
\begin{equation}\label{eq:e8}
\epsilon_{\beta; s,L} = (k_{s,L})^2, \qquad 
k_{s,L}=  {x_{s,L} \over \beta_W},
\end{equation}
where $x_{s,L}$ is the $s$-th zero of the Bessel function 
$J_\nu(k_{s,L}\beta)$, with 
\begin{equation}\label{eq:e9} 
\nu= \left( {L(L+1)\over 3} +{9\over 4}\right)^{1/2}, 
\end{equation}
while the relevant eigenfunctions  are
\begin{equation}\label{eq:e10} 
\xi_{s,L}(\beta) = c_{s,L} \beta^{-3/2} J_\nu(k_{s,L} \beta), 
\end{equation}
where $c_{s,L}$ are  normalization constants.

Eq. (\ref{eq:e5}) is exactly soluble also in the case in which 
$u(\beta)= \beta^2/2$. In this case, to which we are going to refer 
as the X(5)-$\beta^2$ model, the eigenfunctions are \cite{Mosh1555}
\begin{equation}\label{eq:e11}
F_n^L(\beta)= \left[ {2 n!\over \Gamma 
\left(n+a+{5\over 2}\right)}\right]^{1/2} \beta^a L_n^{a+{3\over 2}}(\beta^2)
e^{-\beta^2/2}, 
\end{equation} 
where $\Gamma(n)$ stands for the $\Gamma$-function, $L_n^a(z)$ denotes the 
Laguerre polynomials \cite{AbrSt}, and 
\begin{equation}\label{eq:e12} 
a={1\over 2} \left( -3+\sqrt{9+{4\over 3}L(L+1)}\right), 
\end{equation} 
while the energy eigenvalues are 
\begin{equation}\label{eq:e13} 
E_{n,L}= 2n+a+{5\over 2}= 2n+1 + \sqrt{ {9\over 4}+{ L(L+1)\over 3} }, 
\qquad n=0,1,2,\ldots 
\end{equation}

In the above, $n$ is the usual oscillator quantum number. 
One can see that a formal correspondence between the energy levels of the 
X(5) model and the present X(5)-$\beta^2$ 
model, can be established through the relation 
\begin{equation}\label{eq:e14}
n=s-1.
\end{equation} 
It should be emphasized that Eq. (\ref{eq:e14}) expresses just a formal 
one-to-one correspondence between the states in the two spectra, 
while the origin of the two quantum numbers is different, $s$ labeling the 
order of a zero of a Bessel function and $n$ labeling the number of zeros 
of a Laguerre polynomial. In the present notation, the ground state band 
corresponds to $s=1$ ($n=0$). For the energy states the notation 
$E_{s,L} = E_{n+1,L}$ of Ref. \cite{IacX5} will be kept.  

{\bf 2.2 The $\gamma$-part of the spectrum} 

In the original version of the X(5) model \cite{IacX5} the potential 
$u(\gamma)$ in Eq. (\ref{eq:e6}) is considered as a harmonic oscillator 
potential.   
The energy eigenvalues turn out to be 
\begin{equation} \label{eq:e15} 
E(s,L,n_\gamma,K,M) = E_0 +B (x_{s,L})^2 +A n_{\gamma} + C K^2,
\end{equation} 
where $n_\gamma$ and $K$ come from solving Eq. (\ref{eq:e6}) for $u(\gamma)$ 
being a harmonic oscillator potential 
\begin{equation}\label{eq:e16} 
n_\gamma=0, \\ K=0; \quad n_\gamma=1, \\ K=\pm 2, \quad 
n_\gamma=2, \\ K=0, \pm 4; \quad \dots  
\end{equation}
For $K=0$ one has $L=0$, 2, 4, \dots, while for $K\neq 0$ one obtains 
$L=K$, $K+1$, $K+2$, \dots.

A variation of the X(5) model is considered 
in Ref. \cite{IacTri}, in which $u(\gamma)$ is considered not as a harmonic 
oscillator, but  as an infinite well 
\begin{equation}\label{eq:e16a} 
 u(\gamma) = \left\{ \begin{array}{ll} 0 & \mbox{if $\gamma \leq \gamma_W$} \\
\infty  & \mbox{for $\gamma > \gamma_W$} \end{array} \right. .  
\end{equation} 
In this case the energy eigenvalues are given by 
\begin{equation} \label{eq:e17} 
E(s,L,s',K,M) = A (x_{s,L})^2 + B (x_{s',K})^2 -0.89 A K^2,
\end{equation} 
where $x_{s',K}$ is the $s'$-th zero of the Bessel function 
$J_{\nu'}(k_{s',K} \gamma)$, 
with 
\begin{equation}\label{eq:e18} 
\nu'= {K \over 2} , \qquad k_{s',K}={x_{s',K} \over \gamma_W}, \qquad 
(k_{s',K})^2 = \epsilon_{\gamma;s',K}. 
\end{equation}

In the present X(5)-$\beta^2$ model, one can keep in Eq. (\ref{eq:e6}) for 
$u(\gamma)$ a harmonic oscillator potential, as in the X(5) model.
As a consequence, the full spectrum is given by 
\begin{equation} \label{eq:e19}
E(n,L,n_\gamma,K,M)= E'_0+ B' \left( 2n+1+\sqrt{{L(L+1)\over 3}+{9\over 4}}
\right) + A' n_\gamma + C' K^2,
\end{equation}
which is an analogue of Eq. (\ref{eq:e15}). Eq. (\ref{eq:e16}) and the 
discussion following it remain unchanged. 

Yet another variation of the X(5) model is considered in Ref. \cite{Bijker}. 
In this case, when performing the separation of variables in Eq. (\ref{eq:e1})
by using Eq. (\ref{eq:e3}), one keeps the $4 K^2/3$ term in Eq. (\ref{eq:e5})
instead of Eq. (\ref{eq:e6}). As a result, in Eq. (\ref{eq:e5}) the term 
$L(L+1)-K^2$ appears in the place of $L(L+1)$, and the same substitution 
occurs as a consequence in Eqs. (\ref{eq:e9}), (\ref{eq:e12}), 
(\ref{eq:e13}), and (\ref{eq:e19}). In addition, the term $4 K^2/3$ 
disappears from Eq. (\ref{eq:e6}) and, as a consequence, the term containing
$K^2$ is eliminated in Eqs. (\ref{eq:e15}), (\ref{eq:e17}), and 
(\ref{eq:e19}).  
 
{\bf 2.3 Numerical spectra} 

Numerical results for the $\beta$-parts of the energy spectra (which 
correspond to no excitations in the $\gamma$-variable, i.e. to $n_\gamma=0$) 
of the X(5)-$\beta^2$ and X(5) models are shown in Tables 1 and 2. 
All levels are normalized to the energy of the first excited state, 
$E_{1,2}-E_{1,0}=1.0$, where the notation $E_{s,L}=E_{n+1,L}$ is used. 
The model predictions for these bands are 
parameter independent, up to an overall scale, as seen from 
Eqs. (\ref{eq:e8}), (\ref{eq:e13}).
 This is not the case for bands with $n_\gamma\neq 0$, 
since in this case, as seen from Eqs. (\ref{eq:e15}), (\ref{eq:e19})  
the extra parameters $A$, $C$ and $A'$, $C'$ enter respectively. 
Therefore, in the case of the ($n_\gamma=1$, $K=2$)-band, the energies 
are listed in Table 1 after subtracting from them the relevant $L=2$ bandhead, 
using the same normalization as above. In the case of the ($n_\gamma=1$, 
$K=2$)-band, the conventions of Ref. \cite{Bijker}, described at the end 
of the previous subsection, have been used. The $K=0$ bands are not 
affected by these conventions, anyway. 

A comparison between the spectra of the X(5)-$\beta^2$ and X(5) models, 
given in Tables 1 and 2, leads to the following observations: 

a) The members of the ground state band are characterized by the ratios 
\begin{equation}\label{eq:e20}
R_L= { E_{1,L}-E_{1,0} \over E_{1,2}-E_{1,0} } 
\end{equation}
The $R_4$ ratio within the ground state band 
is 2.646 in the case of X(5)-$\beta^2$, as compared to 2.904 in the case
of X(5). Furthermore, all normalized energy levels within the ground state 
band of X(5)-$\beta^2$ are lower than the corresponding X(5) normalized 
energy levels. The same holds within the $n_\gamma=1$ bands. 
Therefore X(5)-$\beta^2$ corresponds to nuclei 
``less rotational'' than the ones corresponding to X(5).  

b) The location of the bandheads of the various $s$-families is described by
the ratios 
\begin{equation}\label{eq:e21} 
\bar R_s = {E_{s,0}-E_{1,0} \over E_{1,2} -E_{1,0}}. 
\end{equation}
The $\bar R_2$ ratio, related to the position of the bandhead 
of the $s=2$ band, is 3.562 in X(5)-$\beta^2$, while it is 5.649
in X(5). In other words, the $s=2$ bandhead in X(5)-$\beta^2$ lies 
much lower than in X(5). The same holds for all bandheads of $s$-families, 
as seen in Table 2. 

c) The $s=2$ bandhead in X(5)-$\beta^2$ lies almost midway between 
the $4_1^+$ state and the $6_1^+$ state of the ground state band
($E_{1,4}$ and $E_{1,6}$ respectively), while 
in X(5) the $s=2$ bandhead is almost degenerate with the $6_1^+$ state 
($E_{1,6}$) of the ground state band. Indeed, in the case of X(5)-$\beta^2$ 
one has from Eq. (\ref{eq:e20}) that $R_4=2.646$ and 
$R_6=4.507$, their midway being 3.577, as compared 
to 3.562, which is the position of the $s=2$ bandhead.  

A difference between the X(5)-$\beta^2$ and X(5) models can be seen by 
considering the ratios \cite{IacX5} 
\begin{equation}\label{eq:e23} 
R'_s=\frac{E_{s,4}-E_{s,0}}{E_{s,2}-E_{s,0}}. 
\end{equation}
In the $X(5)$ case one obtains the series
\begin{equation}\label{eq:e24}
R'_{s=1,2,3,\ldots}=  2.904, 2.798, 2.754, 2.730, 2.714, \ldots
\end{equation}
In addition the following limit holds 
\begin{equation} \label{eq:e25}
\lim_{s\to \infty}R'_s= 2.646 . 
\end{equation}
In contrast, in the framework of the X(5)-$\beta^2$ model the $R'_s$ ratios
are independent of $s=n+1$
\begin{equation}\label{eq:e26}
{R'}_s^{\rm{osc}}= \frac{\sqrt{\frac{107}{3}}-3}{\sqrt{17}-3} \simeq 2.646. 
\end{equation}
In the case of a simple 5-D harmonic oscillator this ratio would have been 
equal to 2. 

The various ratios are shown in Fig. 2. 
We remark that in the X(5) model the rotational collectivity of 
the bands decreases with increasing $s$ (a fact already mentioned 
in Ref. \cite{IacX5}), while in the X(5)-$\beta^2$ model the rotational 
collectivity 
remains invariant with increasing $n=s-1$. Furthermore, the X(5)-$\beta^2$
constant value of the ${R'}_s^{\rm{osc}}$ ratio is the limiting value of the 
X(5) ${R'}_s$ ratio for $s\to \infty$.  

{\bf 2.4. B(E2) transition rates} 

In nuclear structure it is well known that electromagnetic transition rates 
are quantities sensitive to the details of the underlying microscopic 
structure, as well as to details of the theoretical models, much more 
than the corresponding spectra. It is therefore a must to calculate B(E2) 
ratios (normalized to B(E2:$2_1^+\to 0_1^+$)=100) for the X(5) and 
X(5)-$\beta^2$ models . 

The quadrupole operator has the form \cite{WJ1956} 
\begin{equation}\label{eq:e27}
T^{(E2)}_\mu = t \beta \left[ {\cal D}^{(2)}_{\mu,0}(\theta_i) \cos\gamma  
+{1\over \sqrt{2}} ({\cal D} ^{(2)}_{\mu,2}(\theta_i) + 
{\cal D}^{(2)}_{\mu,-2}(\theta_i)  ) \sin\gamma
\right],
\end{equation}
where $t$ is a scale factor, while the B(E2) transition rates are given by 
\begin{equation}\label{eq:e28} 
 B(E2; L_s\to L'_{s'})= {| \langle L_s || T^{(E2)}  || L'_{s'} \rangle |^2 
\over 2L_s+1}.
\end{equation}
The matrix elements of the quadrupole  operator 
involve an integral over the Euler angles, which is the same as in Ref. 
\cite{IacX5} and is performed by using the properties of the Wigner 
${\cal D}$ functions, of which only ${\cal D}_{\mu,0}^{(2)}$ participates, 
since $\gamma\simeq 0$ in Eq. (\ref{eq:e27}) (as mentioned before Eq. 
(\ref{eq:e3})), as well as an integral over $\beta$. After performing 
the integrations over the angles one is left with 
\begin{equation}\label{eq:e41}
B(E2; L_s \to L'_{s'}) = (L_s 2 L'_{s'} | 000)^2  I^2 _{s,L; s', L'}, 
\end{equation}
where the Clebsch--Gordan coefficient $(L_s 2 L'_{s'}| 000)$ appears, 
which determines the relevant selection rules. 
In the case of X(5) the integral over $\beta$ is 
\begin{equation}\label{eq:e42}
I_{s,L; s', L'}=\int \beta \xi_{s,L}(\beta) \xi_{s',L'}(\beta) \beta^4 d\beta,
\end{equation}
which, as seen from Eq. (\ref{eq:e10}),  involves Bessel functions,
while in the case of X(5)-$\beta^2$ the integral has the form 
\begin{equation}\label{eq:e43}
I_{s,L; s', L'} = \int \beta F^L_n(\beta) F^{L'}_{n'} \beta^4 d\beta, 
\end{equation}  
with $n=s-1$ and $n'=s'-1$, which 
involves Laguerre polynomials, as seen from Eq. (\ref{eq:e11}). 

The results for intraband transitions are reported in Table 3, while 
interband transitions are listed in Table 4. All transitions are 
normalized to ${\rm B(E2}: 2_1^+ \to 0_1^+)=100$. The following observations 
can be made: 

a) The ratio of the lowest B(E2)s within the ground state band 
\begin{equation}\label{eq:e29}  
R_{4\to 2} = {{\rm B(E2}:4_1^+ \to 2_1^+) \over {\rm B(E2}: 2_1^+\to 0_1^+)} 
\end{equation}
is 1.7790
 in X(5)-$\beta^2$, while it is 1.5989 in X(5). In general, the normalized 
intraband B(E2)s in X(5)-$\beta^2$ are higher than the corresponding 
normalized B(E2)s in X(5). This is consistent with the fact that
the various bands in X(5)-$\beta^2$ appear to be ``less rotational''
than the corresponding bands in X(5), as remarked above. It is well known 
from experimental data that in near-rotational nuclei the B(E2)s 
within the ground state band 
have high values which increase relatively slowly with increasing initial  
angular momentum, while in near-vibrational nuclei the B(E2)s within the 
ground state band have low values which increase rapidly with inreasing initial
angular momentum (in the absence of bandcrossings). This experimental 
picture is consistent with the intraband B(E2)s listed in Table 3. 

b) As far as interband transitions are concerned, it is seen in Table 4 
that transitions which are strong in X(5) appear also to be strong in 
X(5)-$\beta^2$, while transitions weak in X(5) are weak in 
X(5)-$\beta^2$ as well. 

{\bf 3. A sequence of potentials lying between U(5) and X(5)} 

{\bf  3.1 General} 

The two cases mentioned in the previous section are the only ones 
in which Eq. (\ref{eq:e5}) is exactly soluble, giving spectra characterized 
by $R_4$ ratios 2.646 and 2.904 for X(5)-$\beta^2$ and X(5) respectively. 
However, the numerical solution of Eq. (\ref{eq:e5}) for other potentials 
is a straightforward task. The potentials to be used in Eq. (\ref{eq:e5}) 
have to obey the restrictions imposed by the 24 transformations mentioned 
in \cite{Bohr} and listed explicitly in \cite{Corr}. 

A particularly interesting sequence of potentials is given by 
\begin{equation}\label{eq:30}
u_{2n}(\beta) = {\beta^{2n} \over 2},
\end{equation}  
with $n$ being an integer. For $n=1$ the X(5)-$\beta^2$ case is obtained, 
while for $n \to\infty$ the infinite well of X(5) is obtained
\cite{Bender}, as shown in Fig. 1.   
Therefore this sequence of potentials interpolates 
between the X(5)-$\beta^2$ model and the X(5) model, in the region lying 
between U(5) and X(5).  

{\bf 3.2  Spectra} 

Numerical results for the spectra of the $\beta^4$, $\beta^6$, and $\beta^8$ 
potentials have been obtained through two different methods. In one 
approach, the representation of the position and momentum operators in matrix 
form \cite{Korsch} has been used, while in the other the direct integration 
method \cite{Scheid} has been applied. In the latter, the differential 
equation is solved for each value of $L$ separately, the successive 
eigenvalues for each value of $L$ labeled by $s=1,2,3,\ldots$ (or, 
equivalently, by $n=0,1,2,\ldots$). The two methods give results 
mutually consistent, the second one appearing of more general applicability. 
The results are shown in Tables 1 and 2, where excitation energies relative 
to the ground state, normalized to the excitation energy of the first excited 
state, are exhibited. 
 
In Tables 1 and 2 the model labels X(5)-$\beta^4$, X(5)-$\beta^6$, 
X(5)-$\beta^8$ 
have been used for the above-mentioned potentials, their meaning being that 
the X(5)-$\beta^{2n}$ model 
corresponds to the potential $\beta^{2n}/2$ plugged in 
the differential equation of Eq. (\ref{eq:e5}) 
obtained in the framework of the X(5) model. 
In this notation X(5)-$\beta^{2n}$ with $n\to \infty$ is simply 
the original X(5) model \cite{IacX5}.

From Tables 1 and 2 it is clear that in all bands and for all values of the 
angular momentum, $L$, the potentials $\beta^4$, $\beta^6$, $\beta^8$ 
gradually lead from the X(5)-$\beta^2$  case to the X(5) results in a 
smooth way. The same conclusion is drawn from Fig. 3(a), where several 
levels of the ground state band of each model are shown vs. the angular 
momentum $L$, as well as from Fig. 3(b), where the bandheads of several 
excited bands are shown for each model as a function of the index $s$.        

{\bf 3.3. B(E2) transition rates} 

The calculation of the B(E2)s follows the steps described in subsection 2.4. 
Eq. (\ref{eq:e41}) is still valid, the only difference being that in 
the integral over $\beta$ the wave functions in the present cases are 
known only in numerical form and not in analytic form as in Eqs. 
(\ref{eq:e42}), (\ref{eq:e43}). 

The results of the calculations for intraband transitions are shown 
in Table 3, while interband transitions are shown in Table 4. In addition, 
the normalized B(E2) transition rates within the ground state band of each 
model are shown in Fig. 3(c). In all cases a smooth evolution from 
X(5)-$\beta^2$ to X(5) is seen. Furthermore, the results 
are in agreement to general qualitative expectations: the more rotational  
the nucleus, the less rapid the increase (with increasing initial angular 
momentum) of the B(E2) ratios within the ground state band should be. 
Indeed the most rapid increase is seen in the case of X(5)-$\beta^2$, 
while the slowest increase is observed in the case of X(5). 

{\bf 4. Comparisons to experimental data} 

From the above observations, we conclude that a few key features of the 
X(5)-$\beta^2$ model, which can serve as benchmarks in the search for 
nuclei exhibiting such behavior, are the following: 

a) The $R_4$ ratio (defined in Eq. (\ref{eq:e20})) should be close to 2.646~.

b) The position of the $s=2$ bandhead should be almost midway 
between the $4_1^+$ and $6_1^+$ ($E_{1,4}$ and $E_{1,6}$) states
of the ground state band, the 
$\bar R_2$ ratio (defined in Eq. (\ref{eq:e21}) being 3.562~.

c) The ratio of the lowest B(E2)s within the ground state band, 
$R_{4\to 2}$ (defined in Eq. (\ref{eq:e29})) should be around 1.7790~. 

Analogous remarks can be made in the cases of the X(5)-$\beta^4$, 
X(5)-$\beta^6$, and X(5)-$\beta^8$ models. 

It is clear that the first place to look for nuclei exhibiting 
X(5)-$\beta^{2n}$ behaviour is the region close to nuclei showing 
X(5) structure. The best examples of nuclei corresponding to the X(5) 
structure are so far the $N=90$ isotones $^{152}$Sm \cite{Sm152}, 
$^{150}$Nd \cite{Nd150}, $^{156}$Dy \cite{Dy156}.  
A preliminary search in the rare earths with $N<90$ shows that 
$^{148}$Nd \cite{Nd148} can be a candidate for X(5)-$\beta^2$, $^{158}$Er 
\cite{Er158} can be a candidate for X(5)-$\beta^6$, while $^{160}$Yb 
\cite{Yb160,Sakai} can be a candidate for 
X(5)-$\beta^4$. Existing data for the ground state bands and the 
$\beta_1$-bandheads of these nuclei are compared to the corresponding 
model predictions in Table 5. However, much more detailed information 
on spectra and B(E2) transitions is needed before final conclusions 
can be reached.  

{\bf 5. Conclusion} 

An exactly soluble model, labeled as X(5)-$\beta^2$, has been constructed 
starting from the original Bohr collective Hamiltonian, separating 
the $\beta$ and $\gamma$ variables as in 
the X(5) model of Iachello, and using a harmonic oscillator potential 
for the $\beta$-variable. Furthermore
it has been proved that the potentials $\beta^{2n}$ (with $n$ being integer) 
provide a ``bridge'' between this new X(5)-$\beta^2$ model (occuring for 
$n=1$) and the X(5) model of Iachello (which is obtained by putting 
in the Bohr Hamiltonian an infinite well potential in the 
$\beta$-variable, materialized for $n \to \infty$). 
Parameter-free (up to overall scale factors) predictions for spectra 
and B(E2) transition rates have been given for the potentials $\beta^2$,
$\beta^4$, $\beta^6$, $\beta^8$, called the X(5)-$\beta^2$, X(5)-$\beta^4$, 
X(5)-$\beta^6$, and X(5)-$\beta^8$ models, respectively, lying between 
the U(5) symmetry of the original Bohr Hamiltonian and the X(5) model.  
Hints about nuclei showing this behaviour have been given. 
 
A sequence of potentials interpolating between the U(5) and E(5) 
symmetries should also be worked out. 
Furthermore, one should try to find a sequence of 
potentials interpolating 
between SU(3) and X(5), as well as between O(6) and E(5). In other words, 
one should try to approach E(5) and X(5) ``from the other side''. 
From the classical limit of the O(6) and SU(3) symmetries of the Interacting 
Boson Model \cite{IA} it is clear that for this purpose potentials with a 
minimum at $\beta \neq 0$ should be considered, the Davidson-like  potentials 
\cite{Dav} 
\begin{equation}\label{eq:e31}  
u_{2n}^D(\beta) = \beta^{2n} + {\beta_0^{4n} \over \beta^{2n} }
\end{equation}
being strong candidates. The Davidson potential, corresponding to $n=1$, 
is known to be exactly soluble \cite{Dav,Rowe}. 

Work in these directions is in progress. 

{\bf Acknowledgements} 

The authors are thankful to Rick Casten (Yale), Jean Libert (Orsay), and 
Werner Scheid (Giessen) for illuminating discussions. Support through the NATO 
Collaborative Linkage Grant PST.CLG 978799 is gratefully acknowledged. 


\newpage 
\parindent=0pt

\begin{table}

\caption{Spectra of the X(5)-$\beta^4$, X(5)-$\beta^6$, and X(5)-$\beta^8$ 
models, compared to the predictions of the X(5) (Eq. (\ref{eq:e8}))
and X(5)-$\beta^2$ (Eq. (\ref{eq:e13})) models, for some $s=1$ bands. 
See subsections 2.3 and 3.2 for details. For the ($n_\gamma=1$, $K=2$)-band 
the conventions of Ref. \cite{Bijker} have been used, as mentioned in 
subsection 2.2. 
}

\bigskip

\begin{tabular}{l r r r r r r}
\hline
band & L & X(5)-$\beta^2$ & X(5)-$\beta^4$ & X(5)-$\beta^6$ & 
X(5)-$\beta^8$ & X(5) \\
\hline
$s=1$, $n_{\gamma}=0$, $K=0$ & & & & &   &       \\
    & 0  & 0.000 & 0.000 & 0.000 & 0.000 & 0.000 \\
    & 2  & 1.000 & 1.000 & 1.000 & 1.000 & 1.000 \\
    & 4  & 2.646 & 2.769 & 2.824 & 2.852 & 2.904 \\
    & 6  & 4.507 & 4.929 & 5.125 & 5.230 & 5.430 \\
    & 8  & 6.453 & 7.343 & 7.777 & 8.015 & 8.483 \\
    & 10 & 8.438 & 9.954 &10.721 &11.151 &12.027 \\
    & 12 &10.445 &12.729 &13.922 &14.605 &16.041 \\
    & 14 &12.465 &15.647 &17.359 &18.355 &20.514 \\
    & 16 &14.494 &18.694 &21.013 &22.383 &25.437 \\
    & 18 &16.529 &21.858 &24.871 &26.677 &30.804 \\
    & 20 &18.568 &25.132 &28.923 &31.225 &36.611 \\
    & 22 &20.610 &28.506 &33.159 &36.017 &42.853 \\
    & 24 &22.654 &31.976 &37.571 &41.046 &49.528 \\
    & 26 &24.700 &35.536 &42.151 &46.302 &56.633 \\
    & 28 &26.748 &39.182 &46.895 &51.781 &64.166 \\
    & 30 &28.796 &42.909 &51.795 &57.475 &72.124 \\
\hline
$s=1$, $n_\gamma=1$, $K=2$ & & & & & &        \\
    & 2 & 0.000 & 0.000 & 0.000 & 0.000 & 0.000\\  
    & 3 & 0.907 & 0.925 & 0.932 & 0.936 & 0.943\\
    & 4 & 1.863 & 1.948 & 1.986 & 2.005 & 2.040\\
    & 5 & 2.842 & 3.046 & 3.138 & 3.186 & 3.274\\
    & 6 & 3.836 & 4.206 & 4.377 & 4.468 & 4.639\\
    & 7 & 4.839 & 5.420 & 5.694 & 5.842 & 6.127\\
    & 8 & 5.848 & 6.681 & 7.084 & 7.305 & 7.737\\
    & 9 & 6.860 & 7.986 & 8.543 & 8.850 & 9.465\\
    &10 & 7.876 & 9.333 &10.066 &10.476 &11.310\\
\hline 
\end{tabular}
\end{table}

\newpage 

\begin{table}

\caption{Same as Table 1, but for some $s>1$ bands. 
}

\bigskip

\begin{tabular}{l r r r r r r}
\hline
band & L & X(5)-$\beta^2$ & X(5)-$\beta^4$ & X(5)-$\beta^6$ & 
X(5)-$\beta^8$ & X(5) \\
\hline
$s=2$, $n_\gamma =0$, $K=0$  & & & & & &       \\
    & 0 & 3.562 & 4.352 & 4.816 & 5.091 & 5.649\\
    & 2 & 4.562 & 5.602 & 6.232 & 6.619 & 7.450\\
    & 4 & 6.208 & 7.733 & 8.684 & 9.288 &10.689\\
    & 6 & 8.069 &10.248 &11.629 &12.527 &14.751\\
    & 8 &10.014 &12.990 &14.896 &16.154 &19.441\\
    &10 &11.999 &15.901 &18.419 &20.100 &24.687\\
    &12 &14.007 &18.951 &22.168 &24.331 &30.454\\
    &14 &16.027 &22.125 &26.121 &28.827 &36.723\\
    &16 &18.056 &25.409 &30.267 &33.573 &43.481\\
    &18 &20.091 &28.796 &34.594 &38.559 &50.719\\
    &20 &22.129 &32.278 &39.094 &43.777 &58.429\\
\hline
$s=3$, $n_\gamma=0$, $K=0$  & & & & & &        \\
    & 0 & 7.123 & 9.384 &10.823 &11.758 &14.119\\
    & 2 & 8.123 &10.817 &12.562 &13.710 &16.716\\
    & 4 & 9.769 &13.228 &15.520 &17.054 &21.271\\
    & 6 &11.630 &16.032 &19.004 &21.025 &26.832\\
    & 8 &13.576 &19.050 &22.802 &25.385 &33.103\\
    &10 &15.561 &22.221 &26.838 &30.051 &39.979\\
    &12 &17.568 &25.514 &31.079 &34.983 &47.413\\
    &14 &19.589 &28.916 &35.504 &40.161 &55.377\\
    &16 &21.617 &32.416 &40.103 &45.571 &63.856\\
    &18 &23.652 &36.007 &44.866 &51.202 &72.838\\
    &20 &25.691 &39.683 &49.786 &57.047 &82.315\\
\hline
$s=4$, $n_\gamma=0$, $K=0$  & & & & & &        \\
    & 0 &10.685 &14.956 &17.831 &19.781 &25.414\\
    & 2 &11.685 &16.536 &19.842 &22.105 &28.805\\
    & 4 &13.331 &19.177 &23.235 &26.044 &34.669\\
    & 6 &15.192 &22.225 &27.189 &30.667 &41.717\\
    & 8 &17.137 &25.483 &31.458 &35.689 &49.551\\
    &10 &19.123 &28.882 &35.955 &41.009 &58.033\\
    &12 &21.130 &32.394 &40.643 &46.584 &67.100\\
    &14 &23.150 &36.002 &45.501 &52.392 &76.721\\
    &16 &25.179 &39.699 &50.519 &58.419 &86.876\\
    &18 &27.214 &43.478 &55.689 &64.653 &97.552\\
    &20 &29.253 &47.334 &61.003 &71.089 &108.739\\
\hline
\end{tabular}
\end{table}

\newpage 

\begin{table}

\caption{Intraband B(E2) transition rates for the X(5)-$\beta^4$, 
X(5)-$\beta^6$, and X(5)-$\beta^8$ models, compared to the predictions 
of the X(5) and X(5)-$\beta^2$ models. See subsections 2.4 and 3.3 for 
details. 
}

\bigskip

\begin{tabular}{l r r r r r r r}
\hline
band & $(L_s)_i$ & $(L_s)_f$ &X(5)-$\beta^2$ & X(5)-$\beta^4$ 
& X(5)-$\beta^6$ & X(5)-$\beta^8$ 
& X(5) \\
\hline
$(s=1)\to(s=1)$ &   &      &       &       &       &       &       \\
    & $2_1$& $0_1$& 100.00 & 100.00 & 100.00 & 100.00  & 100.00 \\
    & $4_1$& $2_1$& 177.90 & 169.03 & 165.31 & 163.41  & 159.89 \\
    & $6_1$& $4_1$& 255.18 & 226.15 & 214.62 & 208.83  & 198.22\\
    & $8_1$& $6_1$& 337.06 & 279.88 & 258.09 & 247.31  & 227.60\\
    &$10_1$& $8_1$& 421.32 & 330.45 & 297.02 & 280.71  & 250.85\\
    &$12_1$&$10_1$& 506.85 & 378.25 & 332.37 & 310.24  & 269.73\\
    &$14_1$&$12_1$& 593.11 & 423.67 & 364.85 & 336.77  & 285.42\\
    &$16_1$&$14_1$& 679.84 & 467.07 & 395.01 & 360.94  & 298.69\\
    &$18_1$&$16_1$& 766.88 & 508.74 & 423.25 & 383.18  & 310.11\\
    &$20_1$&$18_1$& 854.13 & 548.89 & 449.86 & 403.84  & 320.04\\
    &$22_1$&$20_1$& 941.54 & 587.72 & 475.10 & 423.16  & 328.79\\
    &$24_1$&$22_1$&1029.06 & 625.37 & 499.14 & 441.35  & 336.57\\
    &$26_1$&$24_1$&1116.68 & 661.98 & 522.13 & 458.56  & 343.54\\
    &$28_1$&$26_1$&1204.36 & 697.64 & 544.19 & 474.91  & 349.84\\
    &$30_1$&$28_1$&1292.10 & 732.44 & 565.43 & 490.49  & 355.55\\
\hline
$(s=2) \to (s=2)$ &   &      &       &       &       &       &       \\
    & $2_2$& $0_2$& 155.69 & 121.99 & 106.03 &  97.23  &  79.52 \\
    & $4_2$& $2_2$& 240.30 & 187.73 & 162.89 & 149.05  & 120.02 \\
    & $6_2$& $4_2$& 316.27 & 239.86 & 205.80 & 187.08  & 146.75 \\
    & $8_2$& $6_2$& 397.68 & 290.57 & 245.80 & 221.73  & 169.31 \\
    &$10_2$& $8_2$& 481.90 & 339.23 & 282.84 & 253.23  & 188.55 \\
    &$12_2$&$10_2$& 567.55 & 385.73 & 317.15 & 281.93  & 205.12 \\
    &$14_2$&$12_2$& 653.98 & 430.22 & 349.09 & 308.22  & 219.55 \\
    &$16_2$&$14_2$& 740.88 & 472.91 & 379.00 & 332.49  & 232.24 \\
    &$18_2$&$16_2$& 828.08 & 514.03 & 407.16 & 355.03  & 243.52 \\
    &$20_2$&$18_2$& 915.48 & 553.74 & 433.81 & 376.11  & 253.63 \\ 
\hline 
$(s=3) \to (s=3)$ &   &      &       &       &       &       &       \\
    & $2_3$& $0_3$& 211.85 & 144.41 & 116.82 & 102.55  &  72.52 \\
    & $4_3$& $2_3$& 302.74 & 208.42 & 169.03 & 148.48  & 104.36 \\
    & $6_3$& $4_3$& 377.38 & 256.28 & 206.61 & 180.79  & 124.81 \\
    & $8_3$& $6_3$& 458.35 & 304.07 & 242.92 & 211.42  & 142.94 \\
    &$10_3$& $8_3$& 542.55 & 350.70 & 277.41 & 240.11  & 159.02 \\
    &$12_3$&$10_3$& 628.33 & 395.71 & 309.93 & 266.81  & 173.30 \\
    &$14_3$&$12_3$& 714.93 & 439.06 & 340.58 & 291.71  & 186.06 \\
    &$16_3$&$14_3$& 802.00 & 480.86 & 369.55 & 314.99  & 197.54 \\
    &$18_3$&$16_3$& 889.36 & 521.25 & 397.03 & 336.85  & 207.93 \\
    &$20_3$&$18_3$& 976.92 & 560.37 & 423.18 & 357.46  & 217.41 \\
\hline 
\end{tabular}
\end{table}

\newpage 
\setcounter{table}{2} 

\begin{table}

\caption{(continued)} 

\bigskip

\begin{tabular}{l r r r r r r r}
\hline
band & $(L_s)_i$ & $(L_s)_f$ &X(5)-$\beta^2$ & X(5)-$\beta^4$ 
& X(5)-$\beta^6$ & X(5)-$\beta^8$ 
& X(5) \\
\hline
$(s=4) \to (s=4)$ &   &      &       &       &       &       &       \\
    & $2_4$& $0_4$& 268.23 & 165.90 & 127.76 & 108.86  &  69.06 \\
    & $4_4$& $2_4$& 365.19 & 229.20 & 177.27 & 151.33  &  95.96 \\
    & $6_4$& $4_4$& 438.49 & 273.60 & 210.90 & 179.61  & 112.50 \\
    & $8_4$& $6_4$& 519.04 & 318.87 & 244.18 & 207.11  & 127.62 \\
    &$10_4$& $8_4$& 603.25 & 363.62 & 276.37 & 233.38  & 141.39 \\
    &$12_4$&$10_4$& 689.16 & 407.17 & 307.10 & 258.20  & 153.88 \\
    &$14_4$&$12_4$& 775.94 & 449.36 & 336.35 & 281.61  & 165.22 \\
    &$16_4$&$14_4$& 863.19 & 490.22 & 364.21 & 303.71  & 175.57 \\
    &$18_4$&$16_4$& 950.72 & 529.83 & 390.81 & 324.64  & 185.07 \\
    &$20_4$&$18_4$&1038.42 & 568.29 & 416.27 & 344.52  & 193.82 \\
\hline 
\end{tabular}
\end{table}

\newpage 

\begin{table}

\caption{Same as Table 3, but for interband transitions. 
}

\bigskip

\begin{tabular}{l r r r r r r r}
\hline
band & $(L_s)_i$ & $(L_s)_f$ &X(5)-$\beta^2$ & X(5)-$\beta^4$ 
& X(5)-$\beta^6$ & X(5)-$\beta^8$ 
& X(5) \\
\hline
$(s=2) \to (s=1)$ &   &      &       &       &       &       &       \\
    & $0_2$& $2_1$& 121.92 &  93.21 & 81.03  & 74.66   &  62.41 \\
    & $2_2$& $0_1$&   1.57 &   2.04 &  2.18  &  2.21   &   2.12 \\
    & $2_2$& $2_1$&  13.40 &  11.34 & 10.28  &  9.66   &   8.22\\
    & $2_2$& $4_1$&  96.85 &  65.53 & 53.55  & 47.59   &  36.56\\
    & $4_2$& $2_1$&   0.06 &   0.48 &  0.72  &  0.84   &   0.94\\
    & $4_2$& $4_1$&  12.41 &   9.63 &  8.37  &  7.68   &   6.10\\
    & $4_2$& $6_1$&  96.68 &  59.53 & 46.23  & 39.78   &  27.87\\
    & $6_2$& $4_1$&   0.03 &   0.16 &  0.37  &  0.49   &   0.64\\
    & $6_2$& $6_1$&  12.32 &   8.84 &  7.41  &  6.64   &   4.92\\
    & $6_2$& $8_1$&  95.89 &  54.68 & 40.71  & 34.09   &  21.85\\
    & $8_2$& $6_1$&   0.12 &   0.08 &  0.27  &  0.39   &   0.56\\
    & $8_2$& $8_1$&  12.34 &   8.29 &  6.72  &  5.90   &   4.09\\
    & $8_2$&$10_1$&  95.03 &  50.85 & 36.56  & 29.91   &  17.64\\
    &$10_2$& $8_1$&   0.19 &   0.05 &  0.23  &  0.35   &   0.52\\
    &$10_2$&$10_1$&  12.37 &   7.86 &  6.19  &  5.35   &   3.49\\
    &$10_2$&$12_1$&  94.30 &  47.79 & 33.36  & 26.76   &  14.61\\
\hline
$(s=3) \to (s=2)$ &   &      &       &       &       &       &       \\
    & $0_3$& $2_2$& 241.37 & 166.55 & 136.53 & 120.61  &  86.33 \\
    & $2_3$& $0_2$&   2.74 &   3.20 &   3.19 &   3.11  &   2.66 \\
    & $2_3$& $2_2$&  25.45 &  19.61 &  16.82 &  15.19  &  11.25 \\
    & $2_3$& $4_2$& 193.64 & 120.83 &  94.54 &  81.36  &  54.01 \\
    & $4_3$& $2_2$&   0.11 &   0.70 &   0.97 &   1.08  &   1.12 \\
    & $4_3$& $4_2$&  23.75 &  17.14 &  14.27 &  12.67  &   8.83 \\
    & $4_3$& $6_2$& 193.35 & 111.85 &  84.29 &  70.99  &  43.76 \\
    & $6_3$& $4_2$&   0.04 &   0.22 &   0.47 &   0.59  &   0.71 \\
    & $6_3$& $6_2$&  23.73 &  16.09 &  13.01 &  11.37  &   7.46 \\
    & $6_3$& $8_2$& 191.71 & 104.04 &  75.89 &  62.68  &  36.03 \\
    & $8_3$& $6_2$&   0.20 &   0.11 &   0.33 &   0.46  &   0.60 \\
    & $8_3$& $8_2$&  23.89 &  15.32 &  12.07 &  10.39  &   6.44 \\
    & $8_3$&$10_2$& 189.99 &  97.61 &  69.18 &  56.16  &  30.26 \\
    &$10_3$& $8_2$&   0.33 &   0.07 &   0.28 &   0.41  &   0.56 \\
    &$10_3$&$10_2$&  24.05 &  14.69 &  11.31 &   9.61  &   5.65 \\
    &$10_3$&$12_2$& 188.51 &  92.33 &  63.80 &  50.99  &  25.87 \\
\hline 
\end{tabular}
\end{table}

\newpage 
\setcounter{table}{3}

\begin{table}

\caption{(continued)}

\bigskip

\begin{tabular}{l r r r r r r r}
\hline
band & $(L_s)_i$ & $(L_s)_f$ &X(5)-$\beta^2$ & X(5)-$\beta^4$ 
& X(5)-$\beta^6$ & X(5)-$\beta^8$ 
& X(5) \\
\hline
$(s=3) \to (s=1)$  &   &      &       &       &       &       &       \\
    & $0_3$& $2_1$& 0.8371 & 0.0300 & 0.0461 & 0.1835  & 0.5852 \\
    & $2_3$& $0_1$& 0.1178 & 0.0311 & 0.0036 & 0.0002  & 0.0213 \\
    & $2_3$& $2_1$& 0.4123 & 0.0770 & 0.0063 & 0.0012  & 0.0546 \\
    & $2_3$& $4_1$& 0.0170 & 0.0716 & 0.2433 & 0.3876  & 0.6769 \\
    & $4_3$& $2_1$& 0.0059 & 0.0005 & 0.0033 & 0.0139  & 0.0605 \\
    & $4_3$& $4_1$& 0.3111 & 0.0471 & 0.0012 & 0.0051  & 0.0733 \\
    & $4_3$& $6_1$& 0.0049 & 0.1241 & 0.2795 & 0.4046  & 0.6616 \\
    & $6_3$& $4_1$& 0.0022 & 0.0020 & 0.0107 & 0.0240  & 0.0790 \\
    & $6_3$& $6_1$& 0.2554 & 0.0323 & 0.0001 & 0.0085  & 0.0866 \\
    & $6_3$& $8_1$& 0.0169 & 0.1235 & 0.2503 & 0.3548  & 0.5907 \\
    & $8_3$& $6_1$& 0.0090 & 0.0043 & 0.0129 & 0.0255  & 0.0833 \\
    & $8_3$& $8_1$& 0.2165 & 0.0236 & 0.0000 & 0.0104  & 0.0930 \\
    & $8_3$&$10_1$& 0.0220 & 0.1102 & 0.2134 & 0.3011  & 0.5207 \\
    &$10_3$& $8_1$& 0.0130 & 0.0051 & 0.0126 & 0.0240  & 0.0824 \\
    &$10_3$&$10_1$& 0.1877 & 0.0181 & 0.0002 & 0.0112  & 0.0949 \\
    &$10_3$&$12_1$& 0.0231 & 0.0960 & 0.1813 & 0.2555  & 0.4610 \\
\hline 
$(s=4) \to (s=3)$ &   &      &       &       &       &       &       \\
    & $0_4$& $2_3$& 359.75 & 228.92 & 179.59 & 154.38  &  99.18 \\
    & $2_4$& $0_3$&   3.77 &   4.05 &   3.86 &   3.66  &   2.85 \\
    & $2_4$& $2_3$&  36.92 &  26.34 &  21.65 &  19.06  &  12.76 \\
    & $2_4$& $4_3$& 290.41 & 169.37 & 127.74 & 107.41  &  64.60 \\
    & $4_4$& $2_3$&   0.14 &   0.84 &   1.12 &   1.22  &   1.17 \\
    & $4_4$& $4_3$&  34.54 &  23.40 &  18.80 &  16.32  &  10.39 \\
    & $4_4$& $6_3$& 290.00 & 158.78 & 116.23 &  96.05  &  54.31 \\
    & $6_4$& $4_3$&   0.06 &   0.26 &   0.52 &   0.64  &   0.73 \\
    & $6_4$& $6_3$&  34.61 &  22.27 &  17.49 &  14.98  &   9.05 \\
    & $6_4$& $8_3$& 287.51 & 149.11 & 106.29 &  86.50  &  46.15 \\
    & $8_4$& $6_3$&   0.27 &   0.12 &   0.36 &   0.49  &   0.60 \\
    & $8_4$& $8_3$&  34.93 &  21.45 &  16.48 &  13.95  &   8.02 \\
    & $8_4$&$10_3$& 284.90 & 140.88 &  98.06 &  78.69  &  39.76 \\
    &$10_4$& $8_3$&   0.45 &   0.08 &   0.31 &   0.43  &   0.56 \\
    &$10_4$&$10_3$&  35.24 &  20.74 &  15.63 &  13.08  &   7.19 \\
    &$10_4$&$12_3$& 282.67 & 133.99 &  91.26 &  72.30  &  34.73 \\
\hline 
\end{tabular}
\end{table}

\newpage

\begin{table}

\caption{Experimental spectra of the ground state (g.s.) and $\beta_1$ bands 
of $^{148}$Nd \cite{Nd148}, $^{160}$Yb \cite{Yb160,Sakai}, and $^{158}$Er 
\cite{Er158}, compared to the predictions of the X(5)-$\beta^2$, 
X(5)-$\beta^4$, and X(5)-$\beta^6$ models respectively.  
}

\bigskip

\begin{tabular}{l r r r r r r r}
\hline
band & $L$ & $^{148}$Nd & X(5)-$\beta^2$ & $^{160}$Yb & X(5)-$\beta^4$ & 
$^{158}$Er & X(5)-$\beta^6$  \\
\hline
 &    &       &       &       &       &       &       \\
g.s.& &       &       &       &       &       &       \\
 & 2  & 1.000 & 1.000 & 1.000 & 1.000 & 1.000 & 1.000 \\
 & 4  & 2.493 & 2.646 & 2.626 & 2.769 & 2.744 & 2.824 \\
 & 6  & 4.242 & 4.507 & 4.718 & 4.929 & 5.050 & 5.125 \\
 & 8  & 6.153 & 6.453 & 7.142 & 7.343 & 7.772 & 7.777 \\
 & 10 & 8.194 & 8.438 & 9.761 & 9.954 &10.786 &10.721 \\
 & 12 &10.298 &10.445 &12.903 &12.729 &13.952 &13.922 \\
 & 14 &       &       &       &       &17.561 &17.359 \\
 &    &       &       &       &       &       &      \\
$\beta_1$ & & &       &       &       &       &       \\
 &  0 & 3.039 & 3.562 & 4.463 & 4.352 & 4.197 & 4.816 \\
\hline
\end{tabular}
\end{table}


\centerline{\bf Figure captions} 

\bigskip
{\bf Fig. 1.} The potentials $\beta^{2n}$, with $n=1$ (harmonic oscillator, 
solid line), $n=2$ (dash line), $n=3$ (dash dot), $n=4$ (dot), $n=8$ 
(das dot dot), $n=16$ (short dash dot), $n=32$ (short dot), gradually 
approaching (with increasing $n$)  the infinite well potential.  

\medskip
{\bf Fig. 2} (Color online) 
The energy ratio $R'_s$, defined in Eq. (\ref{eq:e23}), for the X(5) and 
X(5)-$\beta^2$ models. See subsection 2.3 for further discussion.  

\medskip
{\bf Fig. 3} (Color online) 
(a) Levels of the ground state bands of the models 
X(5)-$\beta^{2n}$ with $n=1$-4 and of the X(5) model, vs. the angular 
momentum $L$. In each model all levels are normalized to the energy 
of the first excited state. See subsection 3.2 for further discussion. 
(b) Bandhead energies of excited bands of the same models and with the 
same normalization, vs. the  band index $s$. See subsection 3.2 for 
further discussion. (c) B(E2:$L_f+2 \to L_f$) transition rates within the 
ground state bands of the same models, vs. the angular momentum of the 
final state, $L_f$. In each model all rates are normalized to the one 
between the lowest states, B(E2:$2\to 0$). See subsection 3.3 for further 
discussion. 

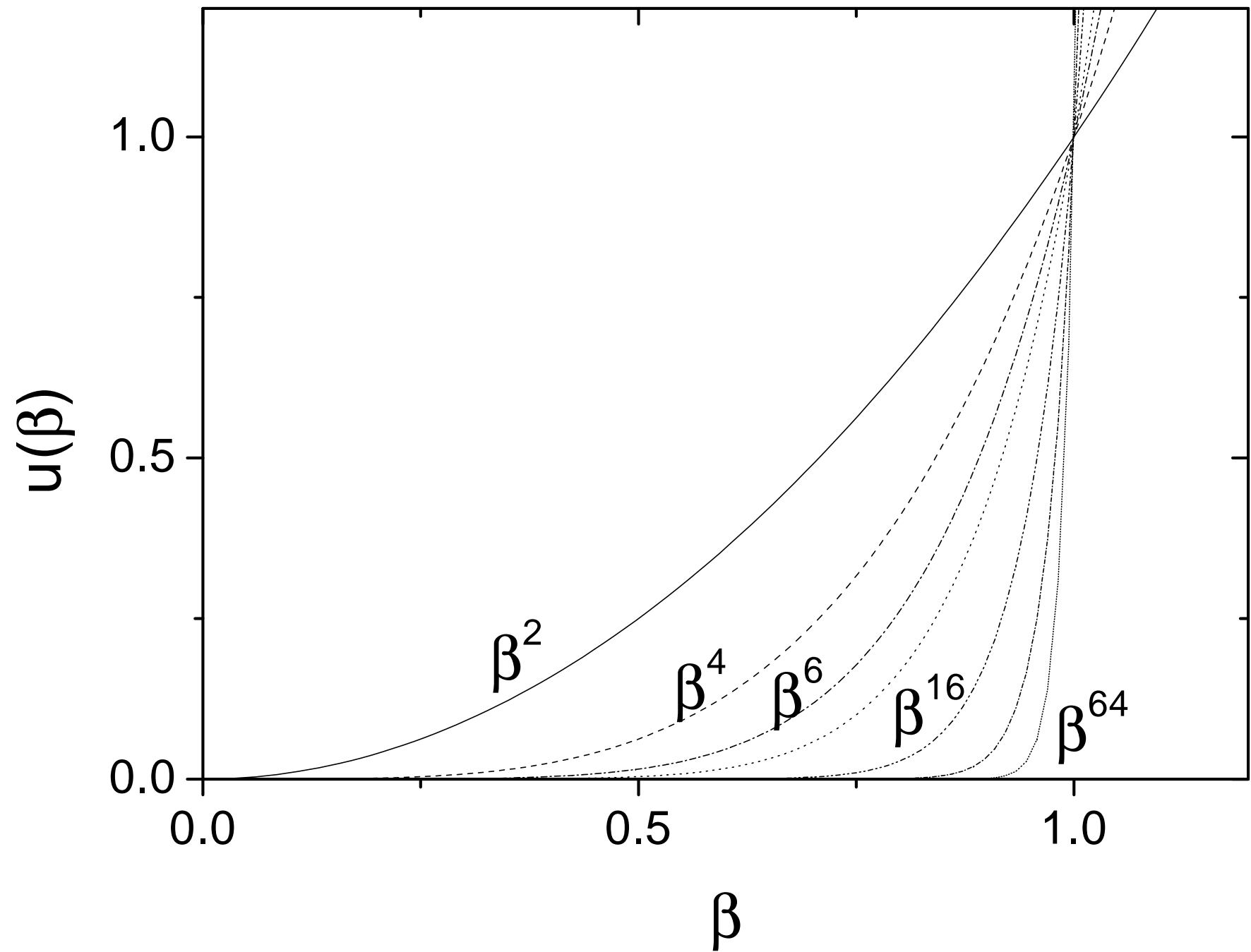
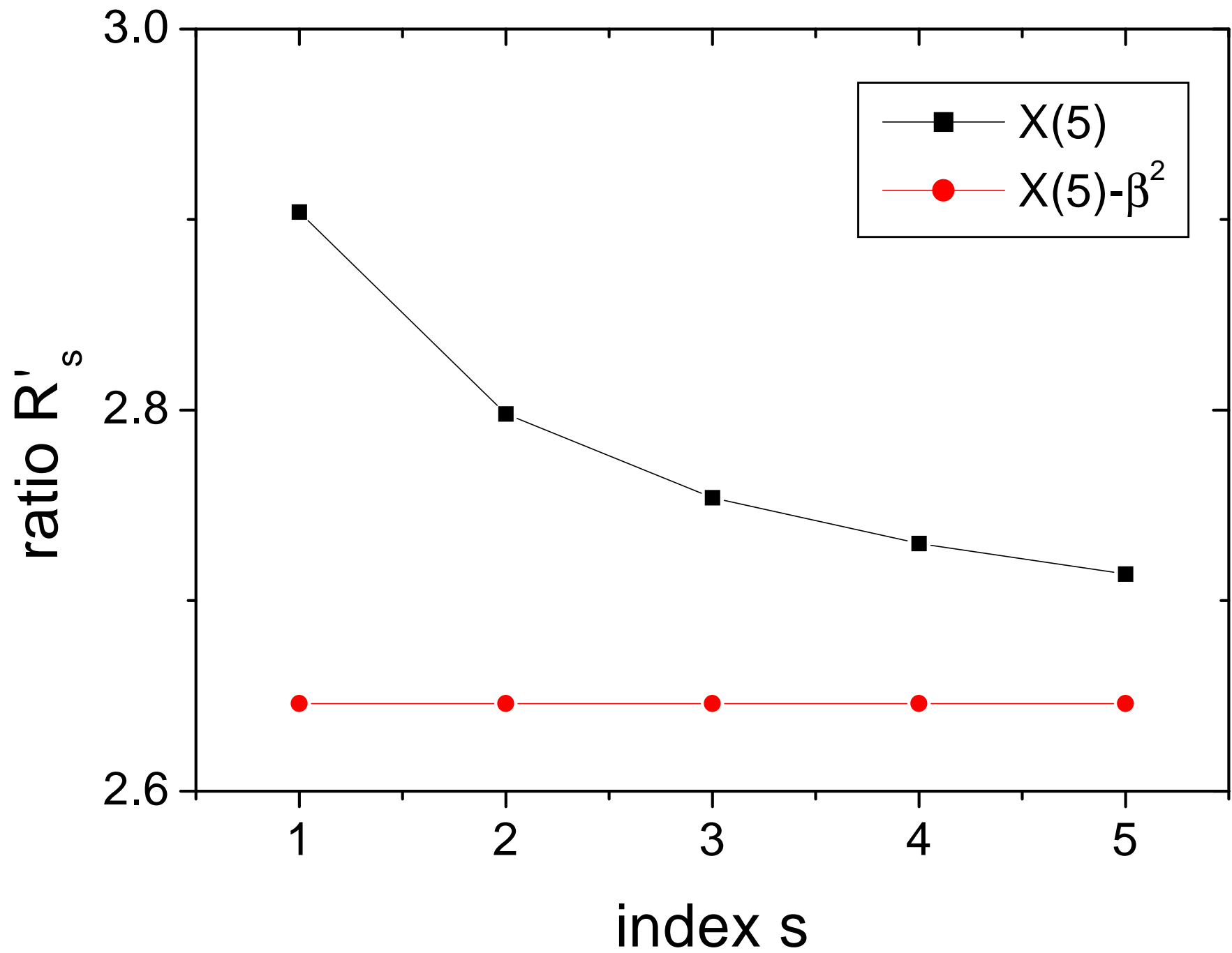
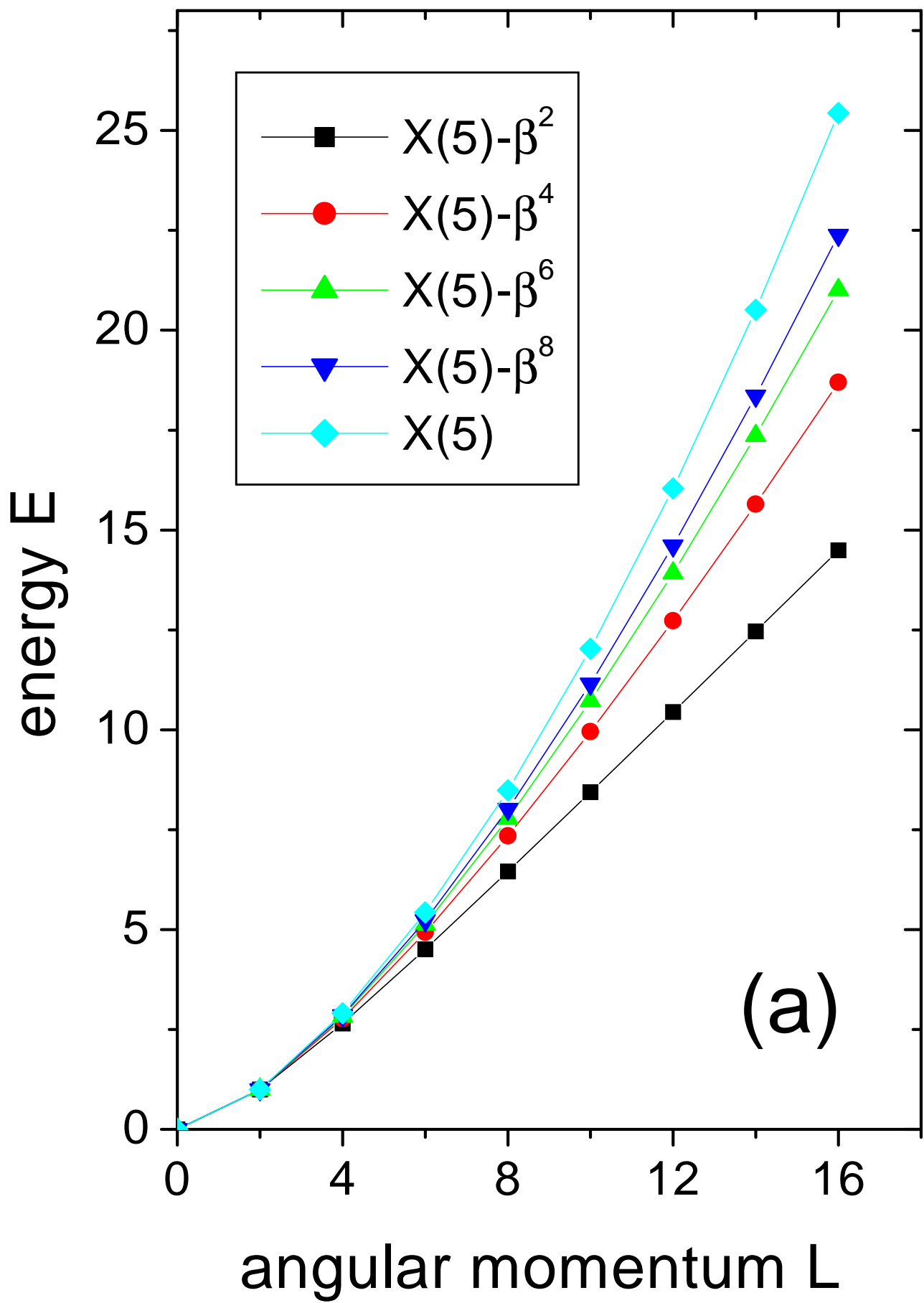
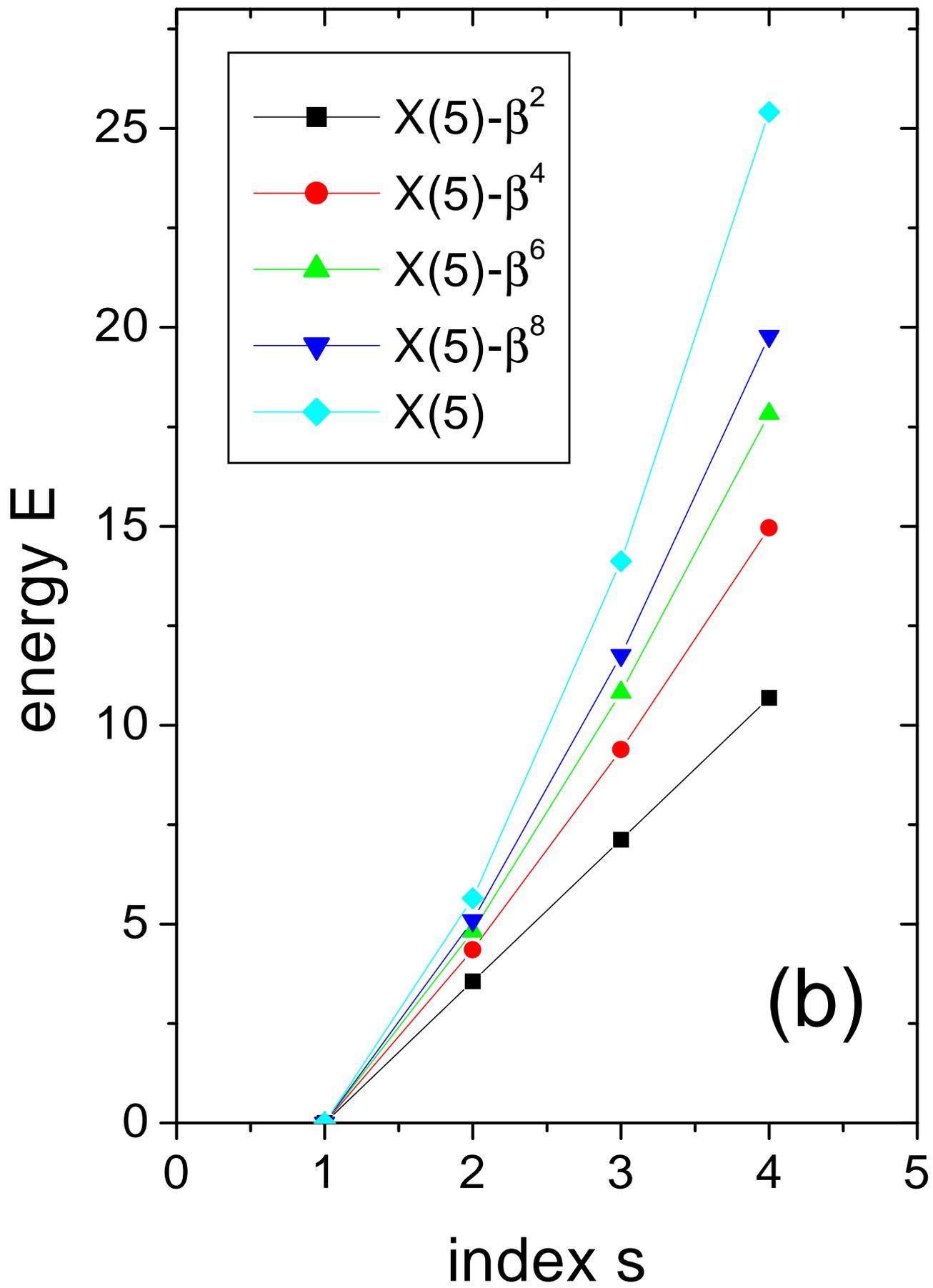
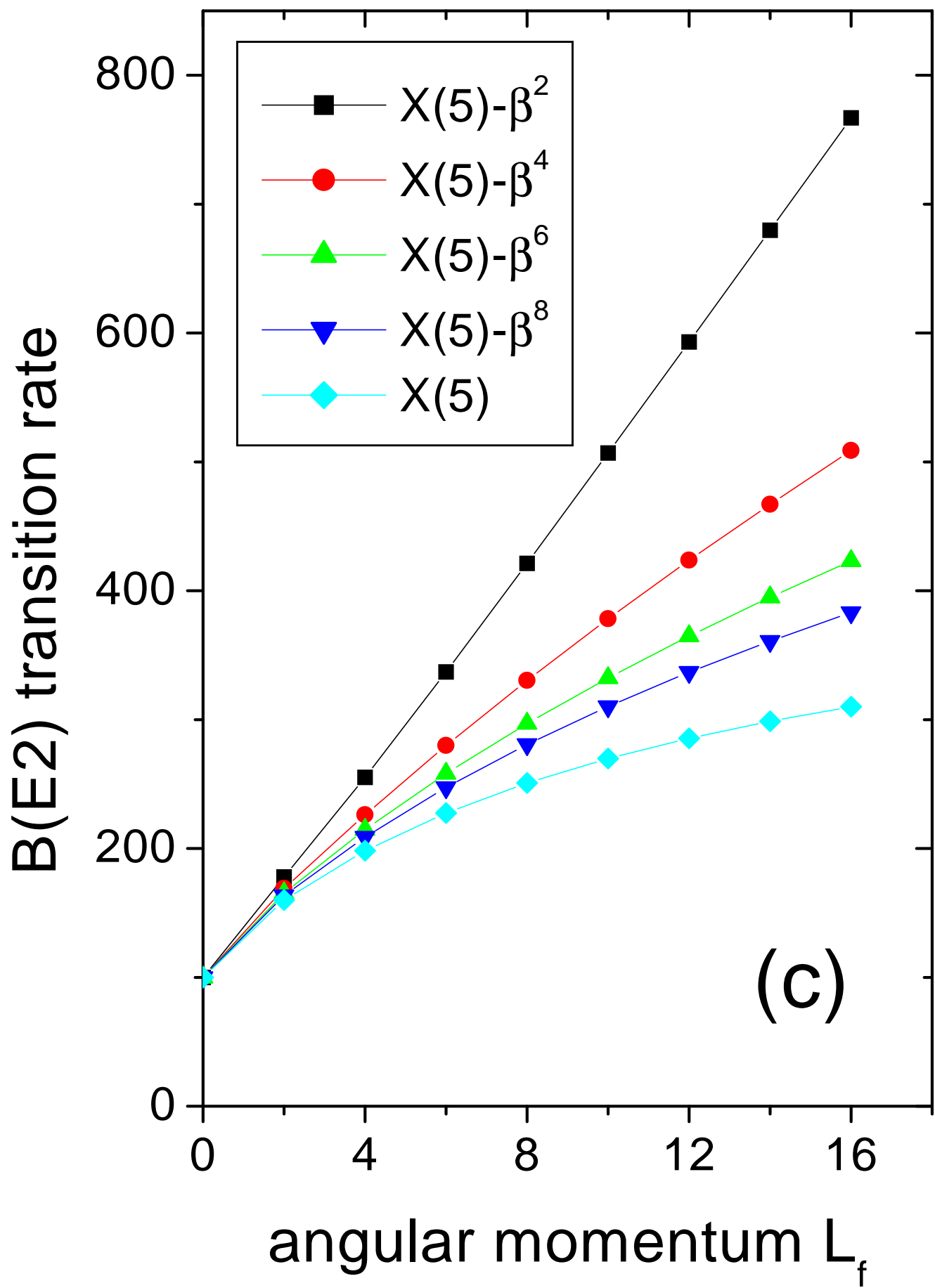
\end{document}